\documentclass[rfc]{article}
\usepackage{epsfig}
\title{A Proposal to Separate Handles from Names\\ on the Internet}
\author{Michael J. O'Donnell\\ \emph{The University of Chicago}}
\date{11 February, 2003}
\begin{document}
\maketitle
\begin{abstract}
  
  Networked communications inherently depend on the ability of the
  sender of a message to indicate through some token how the message
  should be delivered to a particular recipient. The tokens that refer
  messages to recipients are variously known as \emph{routes},
  \emph{addresses}, \emph{handles}, and \emph{names}, ordered by their
  relative nearness to network topology vs.\ human meaning. All four
  sorts of token refer in some way to a recipient, but they are
  controlled by different authorities and their meanings depend on
  different contextual parameters.

  Today's global Internet employs dynamically determined routes, IP
  addresses, and domain names. Domain names combine the functions of
  handles and names. The high value of domain names as names leads to
  substantial social and legal dispute about their assignment,
  degrading their value as handles. The time has come to provide a
  distinct open network handle system (ONHS), using handles that are
  not meaningful in natural language and are therefore not subject to
  the disputes surrounding the use of names.
  
  A handle service may be deployed easily as a handle domain within
  the current Domain Name System. In order to minimize the
  administrative load, and maximize their own autonomy, netizens may
  use public-key cryptography to assign their own handles.
  
\end{abstract}

\section{The Value of Names Leads to Conflict}

The success of the Internet has made it valuable, and that value has
naturally led to conflict between contenders for the profits. Many
parts of the Internet design have avoided such conflict by providing a
sufficient supply of valuable resources, and by economic network
effects that make one party's holdings even more valuable when others
gain similar holdings. The assignment of high-level names in the
Domain Name System (DNS) stands out for the substantial and increasing
contention for scarce resources. There is some controversy over the
extent to which this scarcity should be cured by introduction of a
large number of top-level domains (TLDs). But to some substantial
degree the conflict derives from the natural scarcity of human memory
and attention, which often causes the opposite of a network
effect---other uses of my name can diminish the valuable distinction
of having that name refer to me.

Dispute over valuable names precedes the Internet, and led to legal
regulation of trade marks names. The sudden success of the Internet
threw a monkey wrench into that regulation by changing and blurring
the boundaries between the different contexts in which a name is
used.

Long before highways were super, much less informational, Cyrus Avery
recognized the importance of naming when he promoted the development
of Route 66 through his hometown of Tulsa Oklahoma. Even after the
Road Designation Committee had agreed to construct pavement connecting
Chicago, Tulsa, and Los Angeles, Avery fought to have that route given
a single name, and the highly mnemonic name of Route 60. He settled
for 66, which was at least more mnemonic than 64 or 68. This naming
coup established the notion that people would travel from Chicago to
Los Angeles through Tulsa, enticed song-writers and television
producers to advertise that notion for free, and brought lots of
tourist money to Oklahoma that would otherwise have landed
elsewhere~\cite{route66}. The conflict over Internet domain names is
the natural successor to the conflict over road names, and it will not
be completely resolved as long as domain names have an impact on
commercially valuable behavior.

While a certain amount of conflict over name space is probably
unavoidable, we should remove as much of the value of the Internet as
we can from the scope of that conflict. Domain names in the current
DNS conflate two different sorts of utility:
\begin{itemize}
\item they serve as permanent \emph{handles}, referring to a
  particular agent as its address changes due to mobility or to
  changes in network topology;
\item they serve as mnemonic and guessable \emph{names}, establishing
  a connection between humanly meaningful concepts and network
  agents.
\end{itemize}
Meaningless handles are plentiful, and can be provided promiscuously
without suffering a reverse network effect. Names carry a very
substantial structure of meaning, which adds to their value, but also
makes them much more problematic and in particular attracts
conflict.

So, we should separate these two sorts of utility, and support each
one as well as possible independently of the other. With such
independence, conflict over handles should be insignificantly small.
Conflict over names will continue, but the greater flexibility allowed
by independent use of handles and names may allow more effective
management of names as well. If domain names cease to be the only
available network handles, they may compete more sensibly with other
semantic organization of the Internet, such as \emph{Yahoo} and
\emph{Google}. Bob Frankston wrote a very useful analysis of the way
in which conflict over names leads to violations of their use as
handles, and the need for a ``Safe Haven'' for
handles~\cite{safe_haven}.

Although I am only proposing to separate handles from names, I need to
discuss the other two sorts of network reference
tokens---\emph{routes} and \emph{addresses}---and the parties that use
them.

\section{Parties to Network Operations}

I derive the need for the four different sorts of tokens that may be
used in a network---\emph{routes}, \emph{addresses}, \emph{handles},
and \emph{names}---from the need to accommodate a number of different
sorts of parties to network operations, each one assigned different
authority and responsibility, and each one feeling different
incentives. I derive my list of parties intuitively from my
observations of the global Internet. A different list of parties might
demand a different design.

In this article, an \emph{agent} is any entity in which we care to
invest authority. That includes human beings, offices performed by
human beings, corporations, departments within corporations, groups of
human beings acting somewhat co-operatively, computer programs, and
lots of other possibilities. A document may be thought of as a
relatively passive agent, or the curator of a document may be treated
as a different agent when acting as curator than when performing other
roles.

Parties in network operations include:
\begin{description}
\item[Routers:]{Agents that read information in a message somehow
    indicating the desired recipient, and direct the message in order
    to reach that recipient. Today, the word ``router'' often refers
    to a particular sort of network hardware designed especially to
    serve as a router. But all ``hosts'' in Internet terminology are
    also routers. For the purpose of my conceptual overview, a router
    may also be a particular piece of software running on a computer
    host, or any other identifiable agent that participates in
    routing. A network system normally includes a large number of
    routers.}
\item[Network administration:]{The unique collective agent that
    determines the rules by which a network system operates. Network
    administration is normally a loose organization comprising a
    variety of computers, individuals, and participating organizations
    with different detailed incentives. But to the extent that they
    all co-operate in the interest of effective network operation, I
    regard them as constituting a single collective agent.}
\item[Members:]{All agents that send and receive messages across the
    network.}
\item[Communities:]{Loosely co-operating groups of people.}
\end{description}
A particular agent may participate in the network in more than one of
the roles above, or as a member of a larger agent, but I derive my
observations about reference tokens from the separate actions of
agents in their different roles. Notice again that there are lots of
routers, lots of members, and lots of communities, each constituting
an individual agent. There is only one network administration,
although it is a large collective agent.

\section{Routes, Addresses, Handles, and Names}

I will define these four types of tokens in terms of the operations
that we can perform on them. But you might notice that in most cases
we can translate between the four types of tokens, so mathematically
whatever we can do to one we can do to the others. The real
differences lie in which operations must be very efficient, and in who
has the authority to determine the translations. Translation from
names to handles to addresses to routes is called \emph{resolution}.

\begin{description}
\item[Route:]{A token associated with a message directing
    routers how to deliver that message.}
\item[Address:]{A token associated with a particular target location
    in a network. At any location in the network (not just the target
    location), an address should resolve to a route leading to the
    target location.}
\item[Handle:]{A token associated with an agent participating in a
    network. A handle should resolve to an address at which we may
    communicate with that agent.}
\item[Name:]{A token carrying some humanly understandable meaning. A
    name should resolve to the handle corresponding to its humanly
    understandable meaning.}
\end{description}
I use the word ``should'' advisedly in these definitions. A network
system is intended to support the resolution of names to handles to
addresses to routes in a fashion that satisfies all of the
``should''s. But the requirements for resolution of names is
inherently subjective, so the resolution cannot be perfectly reliable.
As we go down the list from names toward routes, the objective quality
and reliability of resolution improves, but it never becomes
perfect. The success of the global Internet depends critically on our
willingness to work with protocols that \emph{should} produce a
particular result, but that fail occasionally. We have found that it
is better in many cases to guard against the consequences of
occasional failures than to try to prevent them.

In this article, I am trying to show the value of including all four
levels of reference, and all three resolution steps, in the design of
a network. But many network designs reduce the number of levels to
three, two, or even one by omitting levels and/or conflating adjacent
levels. I am not aware of any network design in serious use today that
includes all four levels and keeps them clearly separate. When a level
is missing, resolution just skips over that level to the next lower
one (nearer to routes) that is included in the design.

By composing the different resolution steps, all network reference
tokens resolve down to routes. In many cases, we can also run
resolution backwards (e.g., the current \emph{whois} service maps IP
addresses back to domain names that resolve to them). The relative
ease with which each type of token may be mapped to each other type
makes it hard to keep track of the differences. Whenever the support
for a particular type of token is missing, we tend to use another type
of token to approximate it. For example, the English phrase ``Editor
of the \emph{Journal of Irreproducible Results}'' is a \emph{name}
that also works pretty well outside of the network as a \emph{handle},
referring continuously to the abstract agent who edits \emph{JIR} no
matter how that role passes from person to person and how the people
playing the role move from one address to another. But I regard such a
phrase as essentially a name, since its resolution is mediated by
natural human language, and subject to changes in that language.

In the definitions above, I differentiated addresses, handles, and
names according to the different objects that they should refer to
consistently, even while the routes change. An address should always
refer to the same location, a handle to the same agent, and a name to
the same humanly understandable meaning. The meaning of a name is
clearly subjective, but in fact the notions of location and agent are
also fuzzy. Instead of refining the definitions of location and agent,
which I am pretty sure can never be made satisfactorily objective, I
will distinguish addresses, handles, and names according to where we
invest the authority for their resolutions. In effect, this means that
a network \emph{location} is whatever address resolution defines it to
be, a network \emph{agent} is whatever handle resolution defines it to
be, and \emph{meaning} is whatever natural language defines it to be.

\begin{description}
\item[Address to route:]{Network administration has authority over all
    resolutions of addresses to routes. In practice, the network
    administration as a whole usually delegates some of this authority
    to smaller agents participating in the network administration.}
\item[Handle to address:]{Each member of the network may become owner
    of one or more handles. The owner of a handle has authority over
    all resolutions of that handle to an address. Network
    administration may have authority to assign handles to owners.}
\item[Name to handle:]{Human communities have collective authority over the
    resolution of names to handles.}
\end{description}
The intention in the assignment of authority above is to make the
authority over a resource, the responsibility for acceptable use of
that resource, and the incentive to derive value from the resource,
coincide as much as possible. If I have identified the types of
parties well, and matched the resources to them well, then these
authorities, responsibilities, and incentives will coincide well.

To understand the relations between the three types of resolution,
consider the ways in which each type of resolution can vary. All three
vary over time, to deal with mobility on the network, changes in
jargon, etc. But at a given time, they vary differently according to
context within and without the network.
\begin{description}
\item[Address to route:]{Resolution varies according to starting
  location in the network, so that the ending location is constant.}
\item[Handle to address:]{Resolution does not vary except over
    time.}
\item[Name to handle:]{Resolution varies according to linguistic
    context, local differences in language and culture, and anything
    that affects the way people think.}
\end{description}
Roughly, address$\rightarrow$route resolution varies within the
network, handle$\rightarrow$address resolution doesn't vary at all,
and name$\rightarrow$handle resolution varies outside the network.

\section{The System of Parties and Reference Tokens}

\begin{figure}
\epsfig{file=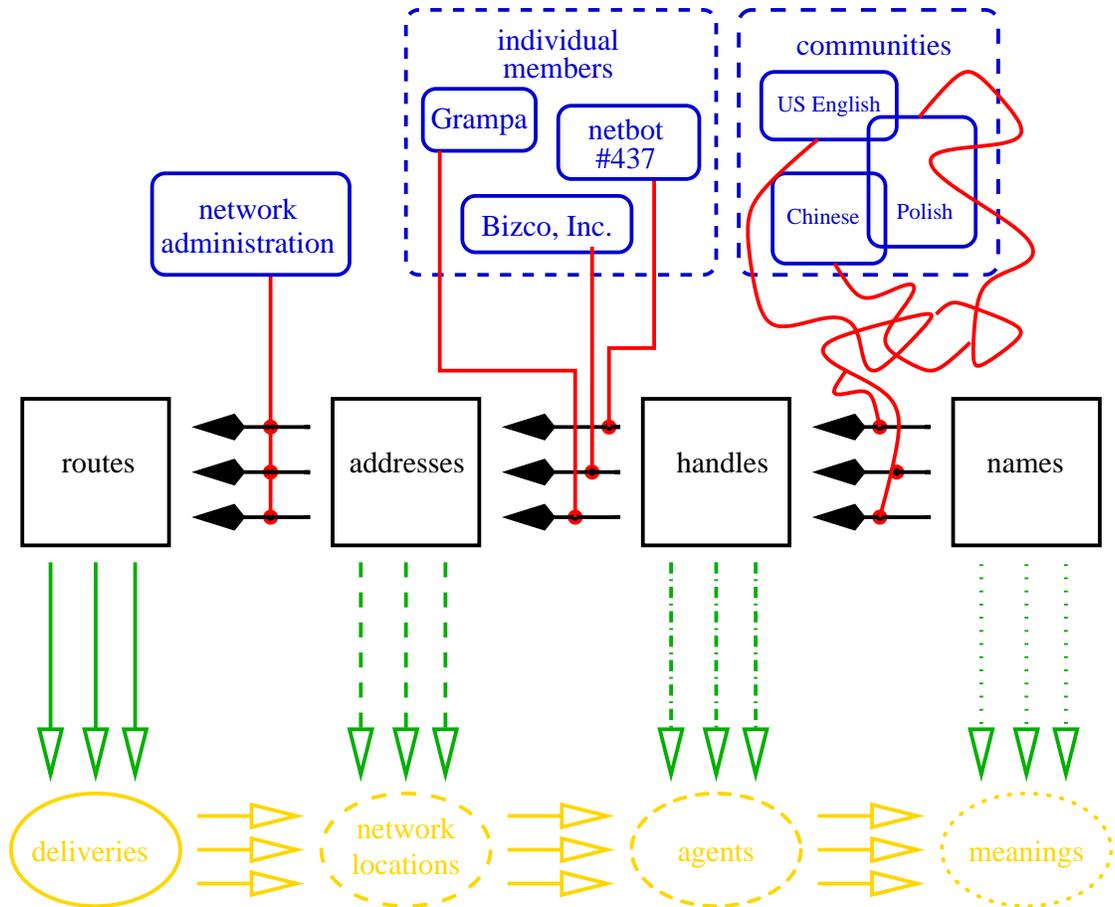}
\caption{\label{fig:system}System of parties,
  reference tokens, and referents.}
\end{figure}

The considerations above suggest a system of parties and reference
tokens with the structure shown in Figure~\ref{fig:system}.
\begin{itemize}
\item{The items in blue rounded rectangular boxes on the top row
    represent the three types of parties with authority over
    resolution---a single network administration, any number of
    individual network members, and any number of overlapping human
    communities. These parties are abstract agents who participate in
    the network, but the network design doesn't define them precisely
    and formally.}
\item{The items in black square boxes in the middle row represent the
    four types of reference tokens. These types of tokens must be
    given explicitly and formally in the network design.}
\item{The items in gold ovals in the bottom row represent the types of
    mental concepts that the four types of reference tokens are
    intended to capture. The act of message delivery in the network is
    rather precisely defined, but moving from left to right, the
    concepts become less objective and more ambiguous, as suggested by
    the fuzzier sorts of boundaries.}
\item{The red lines from the top to the middle row represent the
    authority of the three types of parties to control the three types
    of resolution. The multiple connections from individual members to
    arrows from handles to addresses indicate that each handle owner
    has separate authority over her handle(s).  The squiggly red lines
    from communities to the name$\rightarrow$handle resolution suggest
    the complexity of their collective exercise of authority.}
\item{The black right-left arrows in the middle row represent the
    three types of resolution. These methods of resolution are
    implemented through some sort of tables in various network hosts
    and routers.}
\item{The gold left-right arrows in the bottom row represent the
    conceptual connections that allow each of these concepts to
    determine one to its right. Each delivery leads to a particular
    location. Each location contains a particular agent. Each agent is
    responsible for data or services with a particular human meaning.}
\item{The green up-down arrows between the middle and bottom rows
    represent the intended associations of deliveries with routes,
    locations with addresses, agents with handles, and meanings with
    names. The association of routes with deliveries is determined
    precisely and formally by the network operations. From left to
    right, the associations become less objective and more ambiguous,
    as suggested by the fuzzier sorts of arrows.}
\end{itemize}
The design and implementation of a system of network reference tokens
is intended to make all of the different paths from formal network
tokens (black) to mental concepts (gold) connect the same individual
items. In particular, when we resolve a particular handle into an
address and then into a route, then use that route to deliver a
message to a location, the agent receiving the message at that
location is intended to be the one associated conceptually with the
given handle. As network architects and engineers, we can only control
the mechanisms for the black arrows. A design and implementation are
successful if they make the black resolution arrows work in such a way
that it is possible (with high reliability, but not absolute
perfection) to think up sensible conceptual interpretations of the
green and gold arrows that make these different connections
equivalent.

Notice that there is no fixed definitional foundation in the diagram.
The behavior of the network, as determined by the formal settings of
resolutions in the middle row (black), influences the way that we
think about the concepts in the bottom row (gold). The success of the
system is determined by the utility of its entire behavior, not by the
agreement of one part of the diagram with a completely predetermined
structure in another.

There are a lot of details involved in making such a system of names,
handles, addresses, and routes work efficiently. For example,
although each party should keep a table of the resolutions directly
under its authority, and that table should be the final resort to
resolve tokens correctly, all sorts of routers and other agents should
keep local tables, called \emph{caches}, of the resolutions that they
are using regularly, to save the traffic and the delay associated with
sending to the authoritative source for each resolution. Furthermore,
local caches don't necessarily correspond directly to
address$\rightarrow$route, handle$\rightarrow$address, and
name$\rightarrow$handle resolution. If a particular agent is concerned
with the correspondence between names and addresses, it should cache a
table of the direct resolution of names to addresses, derived by
composing the name$\rightarrow$handle and handle$\rightarrow$address
resolutions. With this sort of transitive caching, the cost of
multilevel resolution is not much more than one-level resolution.

\subsection{Routes, Addresses, and Handlenames in the Current Internet}

\paragraph{Routes.} In the current Internet, routes do not need to be
written down in one place. address$\rightarrow$route resolution
interleaves with the execution of a delivery, so that the route is
implicit in the path by which a message is forwarded. The relationship
between IP addresses and routes is a bit more complicated than this
article suggests, since several different routes to the same address
may be used at the same time, and messages may be broken up en route
and reassembled at the end.

\paragraph{Addresses.} The IP routing protocol is described in a form
that uses IP numbers as addresses. IP numbers are just 32-bit numbers.
But IP numbers are not the only sorts of addresses used in the
Internet. The UDP protocol uses the combination of an IP number and a
port number as an extended address. IP numbers essentially allow a
message to be addressed to an entire computer, called a
\emph{host} (although for technical reasons IP numbers actually refer
to network interfaces). UDP addresses allow a message to be addressed
to a particular \emph{application} running on a particular host, such
as a particular sort of server, or a mail recipient. Other protocols
have other notions of address---the HTTP protocol supporting the World
Wide Web uses URLs as addresses. A URL essentially addresses a
particular file on a particular host.

Networking efficiency sometimes requires addresses of
\emph{distributed locations} for servers requiring the resources of
several hosts. For example multihosted Web servers share the messages
to a particular address among several different hosts. As far as I
know, most addresses for distributed locations on the current Internet
are simulated implicitly through some tricks with routing tables
(IPv6~\cite{ipv6} has addresses for multicast and anycast distributed
locations).  Mobility and intermittent connection call for
time-dependent addresses. For the future, we should open our minds to
the possibility that any sort of instructions for contacting a
particular agent may be thought of as an address.

\paragraph{Handlenames.} The Domain Name System~\cite{dns1,dns2} (DNS)
provides tokens that were mainly designed to serve as handles. But
domain names are usually chosen to be mnemonically valuable sequences
of characters, so they also serve as names. The high value of the
names creates conflicts that degrade the value as handles.

\section{A Pseudohistory of Network Reference}

One way to understand the value of the four layers of reference to the
effective use of a network is to consider a partly fictional, but
realistic, history of network development as it might have happened.

\paragraph{First came routes.}

A network cannot deliver messages without routes. The UUCP system
directed all messages by routes of the form
\texttt{host1!host2}\dots\texttt{!hostn}, describing the entire
sequence of ``hosts'' (acting as routers in my terminology) on the
route.  Each host/router kept its own table of hosts/routers with
which it communicated directly. System administration required little
more than agreeing on the general format for describing routes---all
operational routing details could be handled independently by
hosts/routers.

Routes allowed great support for distributed routing, but they were
not portable. If Sally at the host \texttt{gargoyle} discovered the
route to a great online candy store run by Grampa, and wished to share
it with her friend Paul at \texttt{foghorn}, she could not merely send
the route token to Paul---someone had to translate the route. The only
general and reliable way for Sally and Paul to translate the route was
to append the route between \texttt{foghorn} and \texttt{gargoyle},
which they must have known in order to communicate. This led to nasty
long routes with inefficient forwarding. Even if Sally were selfish,
and kept the candy supply to herself, she had to translate the route
when she moved from \texttt{gargoyle} to \texttt{juniper}. Worse,
selfish Sally could rest immobile at \texttt{gargoyle} and still find
that a change in network topology invalidated her treasured route to
Grampa's candy.

\paragraph{Addresses provide independence of starting point and
  topology.}

In small local networks, with all participants connected rigidly and
directly, routes are pretty much indistinguishable from addresses. The
design of IP for ARPANet and the Internet made useful sense of
globally meaningful addresses in a dynamically changing network with
many different sorts of hops between particular communicating hosts.
Addresses in IP were just 32-bit numbers, but they were usually
written in the form \texttt{n1.n2.n3.n4}, where \texttt{n1},
\texttt{n2}, \texttt{n3}, \texttt{n4} are 8-bit numbers written in
base $10$.

Network administration had authority to assign these numbers, but it
could delegate the authority to assign numbers within a subrange. Each
host/router kept track of its own address, the addresses of
hosts/routers with which it communicated directly, and the direction
in which to forward a message with each possible address. Since
$2^{32}=$4,294,967,296 was somewhat too large to allow each
host/router to store a table with a separate entry for each possible
address, routing tables held entries forwarding all addresses in some
numerical range in the same direction, and providing a default
direction for addresses not in the table.

IP routing provided global addresses, whose meanings would not change
according to the location from which they are used. This allowed Sally
to keep track of Grampa's candy store, and share it with Paul at
will. It takes some thought to make sure that the routing could
work efficiently, but real history shows that it did.

IP routing never required anyone to resolve an address into an
explicitly written token presenting the route. Routes were implicit in
the joint distributed actions of all of the hosts/routers. In effect,
the resolution of an address to a route was interleaved with delivery
according to the route. But routes were still there. Those who really
wanted to write them down could generally get them from the
\texttt{traceroute} program.

With IP addresses to pass around, Sally, Paul, and Grampa were all
quite happy, until the candy store's address changed. The address
changed once because Grampa moved his server from space rented on a
shared computer to his own computer, once because he moved the store
to another state with lower business taxes, and several other times
because network topology changed. Even though IP numerical addresses
were not tied rigidly to routes, they had to be assigned so that the
routing tables could keep information efficiently in terms of a small
number of numerical subranges, forwarding all addresses within such a
subrange in the same direction. Both the candy store's own mobility,
and the requirement to maintain efficient routing through a change in
network topology, prevented Grampa's initial IP address from sticking
reliably to the candy store.

\paragraph{DNS provided handlenames.}

The designers of the Internet realized very early that effective use
of the network required the ability to refer permanently and reliably
to an agent whose address might change. So, they invented \emph{domain
names} of the form
\texttt{bottomdomain.subdomain}\dots\texttt{.topdomain} to
serve as permanent handles. Network administration had authority to
assign domain names, but it could delegate that authority
hierarchically even more flexibly than the authority over IP
addresses. Network administration maintained tables translating
domains to addresses. But only the translation of top-level domains
(\texttt{edu}, \texttt{com}, \texttt{org}, etc.), which appear at the
right-hand end of complete domain names, needed to be available
globally. Each top-level domain name could resolve to the address of a
server keeping tables for that domain only, and so on down the
hierarchy. Furthermore, each individual host could maintain its own
local cache of recently or frequently used domain name translations,
avoiding repeated appeal to the authoritative name servers.

With domain names, Grampa could acquire \texttt{ydnac.com} (to avoid
collision with reality, a certain word is written backwards), and
subdivide the business into \texttt{chocolate.ydnac.com},
\texttt{halvah.ydnac.com}, etc.\ at will. Sally could keep track of
\texttt{ydnac.com}, and perhaps her favorite subdomains, use these
domain names from any host on the network, communicate them to Paul at
will. Furthermore, whenever his own mobility, or a change in network
topology, caused the address of the candy store to change, Grampa
could merely update the entry for \texttt{ydnac.com} in the
appropriate authoritative table, and let it spread around to all of
the local caches.

In the story so far, domain names have served as handles, providing
permanent reference to an agent through changes of address. But
Grampa, and all of his actual and potential customers, got a big bonus
as well. The domain name \texttt{ydnac.com} served as a
humanly-meaningful name. Sally and Paul found \texttt{ydnac.com}
fairly easy to remember, type in their emails to one another, spell
out over the telephone, and even to guess at before they knew of the
candy store or whenever they lost their record of its name. Even had
it been permanent, a numerical IP address would not have been so
convenient.

\paragraph{Conflict about names destroys handles.}

Unfortunately, the very knowledge of the humanly-meaningful semantics
associated with \texttt{ydnac.com}, giving it value as a name, became
incompatible with its function as a handle. A number of larger and
more powerful candy companies, as well as the multinational
corporation \emph{C and Y}, all claimed rights to \texttt{ydnac.com},
and it was taken away from Grampa. The bonus value of
\texttt{ydnac.com} as a name led to administrative action violating
its use as a handle~\cite{safe_haven}.

It is tempting to blame those nasty big companies for stomping on
Grampa, but in fact, humanly meaningful names are inherently subject
to forces beyond the authority of an individual handle owner. Human
meaning, by definition, is determined by human communities. An
individual may determine the human meaning of a name among a circle of
friends who accept his influence. But it is fundamentally infeasible
to keep human meaning in line with the arbitrary exercise of authority
that we would like to invest in the owner of a handle. No matter how
cleverly we assign names to start with, some change in society will
ruin the scheme.

We can invest a lot of effort into improving the fairness with which
conflict over domain names is resolved, and supply more and more
domain names to trade off mnemonic quality against cost. But Grampa's
ownership of \texttt{ydnac.com} is inherently a lucky and
unsustainable windfall, which we cannot provide to everybody who wants
it. Whatever contest we set up, only the winner of that contest may
have it.

\paragraph{Separate handles from names.}

At this point, our story ceases to be history and becomes planning.
If we cannot avoid conflict over network names, perhaps we can at
least provide conflict-free permanent handles. By locating a system of
handles without human meaning at a level of abstraction between names
and addresses, we can provide Sally, Paul, and other lovers of
grandfatherly candy with a permanent token by which they may reach
Grampa as long as he cares to respond to it. We can't help Grampa and
his customers hold on to the wonderful mnemonic value of
\texttt{ydnac.com}, but they can't keep that anyway, and it's better
to keep the handle than to fight a losing battle for the name and keep
nothing. \texttt{ydnac.com} is inherently one of a very small number
of short memorable names that naturally suggest online acquisition of
sweets to all English-speaking Internet users, and we can't give
everyone with an interest in candy full authority over it---we should
expect it to go to the strongest contender.

Without \texttt{ydnac.com}, how will Grampa attract attention to his
business? The same way he always did before his unsustainable
domain-name windfall. Although Grampa's candy handle is opaque and
unmemorable, friends and satisfied customers will pass copies of it
around, using Web browsers and other software that will cater to
users' needs to keep track of unmemorable handles. In a pinch, they
will read it off to one another as, say, a 16-digit hexadecimal number
(similar to a 16-digit credit card number). The handle will appear
behind the scenes in pointers, such as the links in Web pages and
their technical successors. People will keep personal directories that
resolve ``My favorite candy store'' to Grampa's candy handle. Grampa
will advertise in venues that match his natural clientele and
advertising budget, and those venues will associate his handle
\emph{locally} with humanly meaningful words, pictures, and other
tokens, since he can't afford to acquire and defend a global
association. Aggressive indexing services, such as the current
\emph{Yahoo} and \emph{Google}, will organize Grampa's candy handle
into their own presentations of the informational structure of the Web
and its technical successors. And, as long as global domain names
last, Grampa can still choose to fight for \texttt{ydnac.com}, or make
a strategic retreat to \texttt{grampasydnac.com}, or fall back further
to \texttt{grampasydnaconmainstreetintinytownusaearthsolarsystem.com},
or \dots. But I think that, in the long run, he will get more
satisfaction from the alternatives.

\section{Implementing Handles as Domain Names}

Since DNS was designed largely to support network handles, no large
software development or deployment is required to support an open
network handle system. We merely need a nice sponsor to provide a
handle domain within the current system of domain names, such as
\texttt{handleroot.nicesponsor.org}, and provide random-looking
numerical handles promiscuously to all who ask for them. For example,
if we choose to present handles in base 16, a typical handle might
look like \texttt{h0061A38F9A3540B9}. This handle may be implemented
under DNS as \texttt{h0061A38F9A3540B9.handleroot.nicesponsor.org}.
There is no particular need for the root of the handle system to be a
top-level domain in DNS. Handle owners may define subdomains below
their handles with complete freedom, as domain owners do today.

The main administrative burden on the sponsor of an open network
handle system is the authentication of updates to the
handle$\rightarrow$address resolution tables. Because an individual
handle has almost no intrinsic value beyond that created by the owner
at the assigned address, there is no need to identify handle owners
reliably with a person or institution in the real world. It is only
important to minimize accidental and malicious capture of handles
after they are assigned and used. To avoid being a party to disputes,
the sponsor should minimize its contact with handle owners.

The precise design of the authentication mechanism should be allowed
to vary. The sponsor may offer handle owners some choice to trade off
authentication overhead against security, depending on owners'
resources and the value of a particular handle. So, in general a
handle might look like \texttt{h<t><p1>...<pn>}, where \texttt{<t>}
indicates the type of authentication used for handle updates, and
\texttt{<p1>...<pn>} are parameters of that method, including the
unique handle key.

A complete design of ONHS will allow a user to
\begin{itemize}
\item create a new handle;
\item (re)assign an address temporarily to a handle;
\item delegate a handle temporarily to another handle, possibly with a
different owner;
\item cancel a handle irrevocably;
\item transfer a handle irrevocably to another handle, usually with a
different authentication key;
\item mark a handle's security irrevocably as compromised.
\end{itemize}
Typical scenarios using these operations to support creation of handle
hierarchies, conversion to new authentication methods, and transfer of
handles connected with sale of a business, are treated
in~\cite{onhs_dns}.

\section{Self-Assigned Cryptographic Handles}

In order to ease the sponsor's administrative burden and potential
liability as much as possible, ONHS should ideally allow completely
independent self-assignment of handles. Such self-assignment is
feasible in principle, using public-key cryptographic signatures. Any
user may create a public/private key pair, and assign herself a handle
of the form \texttt{h1g5k<key>}, where ``\texttt{h}'' just reminds us
that this is a handle (and satisfies DNS' requirement to start with an
alphabetic character), ``\texttt{1}'' indicates a public-key handle,
``\texttt{g}'' introduces a code for the encryption algorithm,
``\texttt{5}'' is the IANA standard number~\cite{rsa/sha1} for the
RSA/SHA1 signature algorithm. Finally, \texttt{<key>} is some number
of hexadecimal digits from the end of the SHA1~\cite{sha1} hash of an
RSA~\cite{rsa} public key. 16 digits will probably provide plenty of
assurance against accidental or malicious collision of handles. We may
allow individual handle owners to choose the tradeoff between length
and security. Ignoring the fact that I haven't used an actual RSA/SHA1
key, the whole embedding of a self-assigned handle in DNS might look
like \texttt{h1g5k0061A38F9A3540B9.handleroot.nicesponsor.org}.

The security extensions to DNS (DNSSEC)~\cite{dnssec} already
implement almost all of the functions required to support
self-assigned keys based on RSA/SHA1~\cite{onhs_dns}.

In a system of self-assigned handles, there is no need to invest any
authority over resolution in the operator of a name server. A user who
queries a handle may authenticate the handle with the owner at the
resolvent address. Since the identity of the public key is embedded in
the handle itself, as long as the particular key and signature
algorithm are not compromised, no intermediary can defraud a
conscientious querier. The operator of a name server becomes a mere
facilitator of the contact between querier and handle owner, with all
responsibility for the authenticity of the contact left to the parties
most concerned. The system provides a lot of flexibility for
individual handle owners and queriers to contract with operators to
provide a defined level of service, without affecting authenticity of
communications. \cite{onhs_dns} describes potential attacks on a
self-assigned ONHS in more detail.

\subsection{Digression on Public-Key Cryptography}

The embedding of (hash keys for) public cryptographic keys has
occurred to a number of people. The \emph{Open Privacy Initiative}
uses it to create \emph{nyms} as the basis for the accumulation of
reputations~\cite{opi-nym}. Carl M.\ Ellison mentioned the use of
hashed public keys as self-assigned identifiers~\cite{ellison}. Ronald
Rivest and Butler Lampson use the same basic idea in their Simple
Distributed Security Infrastructure project~\cite{sdsi_thesis}. Daniel
J.\ Bernstein pointed out the value of hostnames that match public
keys~\cite{bernstein}. Scott Nelson explained the present application
to ONHS in a private communication.

The development of the idea of signatures as objects is well worth
considering as a tangent. At first, public-key cryptography appears to
solve the digital signature problem. On reflection, we see that it
reduces the digital signature problem to the key management
problem. The secret key appears to be the harder one to manage. But it
only needs to be kept by its owner, who has a substantial incentive to
guard it well, and who may limit the number of secret keys that she
manages. Management of public keys is often the bigger problem. Each
key must be published, and somehow the user of the key must be assured
of the connection between that key and a particular agent. The
dominant idea to establish that connection is a \emph{chain of trust}
in which we all store public keys for a small number of globally
trusted agents, who sign certificates of subagents, who sign
certificates of more subagents, and so on. Such a chain is no stronger
than its weakest link, and the establishment of satisfactory roots is
already very difficult.

I know of no simple solution to the problem of secret-key management.
But we may finesse much of the problem of public-key management by
letting keys themselves be identities. In effect, public-key handles,
nyms, and similar objects do not try to solve the problem of
identification on the network. Rather, they serve as tools for
constructing identities on the network. They link together all of our
transactions with those identities in a reasonably reliable way, and
allow us to develop a sense of confidence (or lack of it) from a
sequence of transactions. We may use contacts outside of the network,
and network communications with agents whom we already know and trust,
to increase our confidence in the owner of a particular handle. But
the handle itself does not provide any authentic identification deeper
than ``the owner of this handle.''

So, we may characterize the value of ONHS with a slogan: it provides
continuity, not authenticity. Mere continuity appears to have modest
value, but when available cheaply and reliably it may be the
foundation on which we construct more valuable qualities, such as
authenticity.

\section{Call to Action}

Problems with robust reference to resources on the Internet are widely
recognized. They are currently being addressed by a variety of
ambitious and potentially valuable projects, including the development
of governance systems for DNS by ICANN, Tim
Berners-Lee's Semantic Web~\cite{semantic_web}, and the INET working
groups on Uniform Resource Identifiers (URI)~\cite{uri} and Uniform
Resource Names (URN)~\cite{urn}. All of these projects are wrestling
directly with human meaning in some form, and all have the potential
to add a lot of value to network operations. But all of them are
dealing with unsolved problems that preclude immediate wide
deployment.

The Permanent URL (PURL) service of OCLC~\cite{purl} offers free
handles in the URL name space, providing an immediate useful solution
to the problem of permanent reference to documents and services that
wander around the Web. They are assigning meaningful names on a
first-come-first-served basis, which exposes them to the same sorts of
dispute that plague DNS if their success attracts the wrong sort of
attention. And the service only applies to URL addresses.

The time is ripe for an open network handle system, operating as a
domain within DNS, to offer meaningless numerical handles freely and
promiscuously to all who ask. Handles that resolve to IP numbers
provide a particularly strategic level of service at the foundation of
Internet addressing. Future expansion to higher-level addresses,
including UDP addresses, URLs, and more, is valuable, but should not
delay the implementation of resolution to IP numbers.

The cost of operating the name server will be comparable to the cost
of operating a single DNS root server. If the service is successful,
there will be a sufficient incentive for the community to provide
additional servers to share the load.

If the quality of implementations of DNSSEC is deemed sufficient, the
ONHS should support RSA/SHA1 self-assigned handles. For users who are
not ready to deal with private-key management, there should be a
password-authenticated procsy service to generate and hold private
keys. It is also possible to provide a service assigning random
handles with only password authentication. This should be the fall-back
if DNSSEC implementation is not sufficiently advanced, but the sponsor
of ONHS service should encourage migration to owner-managed keys as
soon as possible, to avoid legal responsibility for keys.

\end{document}